\documentclass[useAMS,usenatbib]{mn2e}

\newcommand\article{{\it Article}}
\newcommand\nd{$\cdots$}
\newcommand\ddofo{DDO$_{51}$}

\usepackage{color}

\newcommand{\acronym}[1]{{\small{#1}}}
\newcommand{\project}[1]{\textsl{#1}}
\newcommand{\apogee}{\acronym{\project{APOGEE}}}
\newcommand{\kepler}{\project{Kepler}}
\newcommand{\wise}{\acronym{\project{WISE}}}
\newcommand{\allwise}{\acronym{\project{ALLWISE}}}
\newcommand{\kic}{\acronym{\project{KIC}}}
\newcommand{\irsa}{\acronym{\project{IRSA}}}
\newcommand{\lamost}{\acronym{\project{LAMOST}}}

\newcommand{\gaia}{\project{Gaia}}

\usepackage{amssymb}
\usepackage{graphicx}

\title{Infrared colours and inferred masses of metal-poor giant stars in the \kepler\ field}
\author[Casey et al]{A.~R. Casey$^{1,2,3}$\thanks{andrew.casey@monash.edu}, G.~M. Kennedy$^{4}$, T.~R. Hartle$^3$, Kevin C. Schlaufman$^5$\\
$^1$School of Physics \& Astronomy, Monash University, Clayton 3800, Victoria, Australia\\
$^2$Faculty of Information Technology, Monash University, Clayton 3800, Victoria, Australia\\
$^3$Institute of Astronomy, University of Cambridge, Madingley Road, Cambridge CB3 0HA, UK\\
$^4$Department of Physics, University of Warwick, Gibbet Hill Road, Coventry, CV4 7AL, UK\\
$^5$Department of Physics and Astronomy, Johns Hopkins University, 3400 N Charles St, Baltimore, MD 21218, USA\\
}
\begin{document}

\def\aj{AJ}                   
\def\araa{ARA\&A}             
\def\apj{ApJ}                 
\def\apjl{ApJ}                
\def\apjs{ApJS}               
\def\ao{Appl.Optics}          
\def\apss{Ap\&SS}             
\def\aap{A\&A}                
\def\aapr{A\&A~Rev.}          
\def\aaps{A\&AS}              
\def\azh{AZh}                 
\def\baas{BAAS}               
\def\jrasc{JRASC}             
\def\memras{MmRAS}            
\def\mnras{MNRAS}             
\def\pra{Phys.Rev.A}          
\def\prb{Phys.Rev.B}          
\def\prc{Phys.Rev.C}          
\def\prd{Phys.Rev.D}          
\def\prl{Phys.Rev.Lett}       
\def\pasp{PASP}               
\def\pasj{PASJ}               
\def\qjras{QJRAS}             
\def\skytel{S\&T}             
\def\solphys{Solar~Phys.}     
\def\sovast{Soviet~Ast.}      
\def\ssr{Space~Sci.Rev.}      
\def\zap{ZAp}                 
\let\astap=\aap
\let\apjlett=\apjl
\let\apjsupp=\apjs

%Accepted 2018 April 30. Received 2018 April 30; in original form 2017 July 14
\date{Submitted 2017 July 14. Accepted 2018 April 30. Received 2018 April 30; in original form 2017 July 14}

\pagerange{\pageref{firstpage}--\pageref{lastpage}} \pubyear{2018}

\maketitle

\label{firstpage}

\begin{abstract}
Intrinsically luminous giant stars in the Milky Way are the only potential
volume-complete tracers of the distant disk, bulge, and halo.  
The chemical abundances of 
metal-poor giants also reflect the compositions of the earliest
star-forming regions, providing the initial conditions for the chemical
evolution of the Galaxy.  However, the intrinsic rarity of metal-poor
giants combined with the difficulty of efficiently identifying them with
broad-band optical photometry has made it difficult to exploit them for
studies of the Milky Way.  One long-standing problem is that photometric
selections for giant and/or metal-poor stars frequently include a large
fraction of metal-rich dwarf contaminants.  We re-derive a giant star
photometric selection using existing public $g$-band and narrow-band
\ddofo\ photometry obtained in the \kepler\ field.  Our selection
is simple and yields a contamination rate of main-sequence stars of $\lesssim$1\% and a completeness of about 80\,\% for giant stars
with $T_{\rm eff} \lesssim 5250\,{\rm K}$ --  subject to the selection function
of the spectroscopic surveys used to estimate these rates, and the magnitude
range considered ($11 \lesssim g \lesssim 15$).
While the \ddofo\
filter is known to be sensitive to stellar surface gravity, we further
show that the mid-infrared colours of \ddofo-selected giants are
strongly correlated with spectroscopic metallicity.  This extends
the infrared metal-poor selection developed by Schlaufman \& Casey,
demonstrating that the principal contaminants in their selection can
be efficiently removed by the photometric separation of dwarfs and
giants. This implies that any similarly efficient dwarf/giant discriminant
(e.g., \gaia\ parallaxes) can be used in conjunction with \wise\ colours
to select samples of giant stars with high completeness and low contamination.
We employ our photometric
selection to identify three metal-poor giant candidates in the \kepler\
field with global asteroseismic parameters and find that masses inferred
for these three stars using standard asteroseismic scaling relations are
systematically over-estimated by 20--175\%.  Taken at face value, this
small sample size implies that standard asteroseismic scaling relations 
over-predict stellar masses for metal-poor giant stars. 
\end{abstract}
%This work shows that existing \ddofo\, $g$, and mid-infrared
%photometry can be used produce accurate and precise estimates of the
%stellar parameters $T_{\rm eff}$, $\log{g}$, and [Fe/H].  


\section{Introduction}
\label{sec:introduction}

Red giant stars are effective tracers of the disk, bulge, and halo
of the Milky Way.  They are especially important for penetrating the
most extincted regions of the bulge \citep[e.g.,][]{Rich_1990,Rich_2007,Casey_Schlaufman_2015}.
In short, they allow for a thorough and relatively unbiased examination
of all major components of the Milky Way and its satellite systems.
Given their importance for tracing the structure and evolution of the
Milky Way, an efficient photometric selection for giant stars has been
long-standing goal of Galactic astronomers.

\citet{Geisler_1984} presented what has become the most influential
photometric dwarf/giant separation.  Geisler's classification of the
narrow-band Washington \citep{Canterna_1976} \ddofo\ filter demonstrated
that the $g-$\ddofo\ colour was significantly different for FGK-type
dwarfs and giants.  The colour separation is the result of strong stellar
absorption features that appear in the spectra of FGK-type stars near
the central wavelength of the \ddofo\ filter.  Specifically, the Mg~I
triplet near 517 nm is one of the strongest spectral features in late-type
stars, with dwarfs showing extended wings induced by pressure broadening.
In addition to the Mg~I triplet lines contributing $\log{g}$ sensitivity
to the \ddofo\ filter, several bands of the MgH $A^2\Pi-X^2\Sigma$
structure are present in this narrow wavelength range.  Despite a 
small secondary dependence on metallicity \citep{Paltoglou_1994,Majewski_2000},
these features are much weaker in giants than dwarfs at the same effective
temperature.  Although the $g$-band filter is broader than the narrow
\ddofo\ filter by more than 100~nm, both responses curves peak near
the same central wavelength.  For these reasons, the $g-$\ddofo\ colour
provides outstanding photometric sensitivity in surface gravity in 
late-type stars.  For comparison, metal-rich main-sequence stars overlap
with very metal-poor giant stars in the $c_{1.0}$--$(b-y)_0$ plane of
Str\"omgren photometry, and giant/dwarf separation using 
Str\"omgren photometry is extremely sensitive to reddening \citep{Arnadottir_2010}.
Thus, the $g-$\ddofo\ colour is among the most promising for distinguishing
giant stars, particularly metal-poor giant stars, from main-sequence stars.

While the \ddofo\ filter is known for its ability to distinguish dwarfs
from giants, it has not seen extensive use in large-scale studies of
Milky Way \citep[though see][]{Majewski_2000,Morrison_2001,Helmi_2003,
Munoz_2005,Saha_2010,Janesh_2016,Slater_2016,Blanton_2017}.
Most galactic studies seeking to assemble a clean (i.e., relatively
uncontaminated) sample of giant stars have focused on later-type M stars.
In constructing the standard $JHKLM$ system, \citet{Bessell_Brett_1988}
showed that dwarfs and giants bifurcated in infrared colours at spectral
types later than M.  This feature allows for a clean sample of either
cool M dwarfs or giants to be easily constructed without the need for
narrow-band photometry.  Given that M giants are more luminous than
FGK type stars, many studies have produced inferences about Milky Way
structure through uncontaminated samples of M giants 
\citep[e.g., see][]{Sheffield_2014,Koposov_2015,Li_2016}.

While it is tempting to assert that M giants selected from public
infrared photometry are sufficient tracers of the Milky Way, it is
well understood that M giants preferentially trace metal-rich stellar
populations.  For this reason, most of the structure in the Milky Way's
metal-poor halo will not appear in even the cleanest M-giant sample.
This can result in biased inferences, leaving the largest component of
the Milky Way (by volume) less than fully understood.

This \article\ is organised in the following manner.  In Section
\ref{sec:data} we use public photometry and spectral data available
in the \kepler\ field to re-derive a giant photometric selection with
high completeness and negligible contamination.  We show that
the stars in the resulting photometrically-selected giant sample have
spectroscopic metallicities that are strongly correlated with infrared
colours.  Confident in our photometric  selection, we then use 
photometrically-selected metal-poor giant stars with
publicly available asteroseismic parameters $\Delta\nu$ and $\nu_{max}$
to show that masses inferred using scaling relations are systematically
over-estimated for metal-poor stars.  This observation suggests that
a modification to the asteroseismic scaling relations is warranted in
the metal-poor regime.  In Section \ref{sec:discussion} we discuss how
these results extend existing work on searches for metal-poor stars using
mid-infrared photometry and the implications for asteroseismic scaling
relations.  Our conclusions follow in Section \ref{sec:conclusions}.

\section{Data \& Analysis}
\label{sec:data}

\subsection{Photometric Selection}
\label{sec:photometric-selection}

% Good to remove "bad" WISE data, keeping w1rchi2+w2rchi2<5 and w1sigmpro+w2sigmpro<0.045 helps

We constructed our photometric selection using the extensive
public photometric and spectroscopic data available in the \kepler\
field.  We first cross matched the \textit{Kepler Input Catalogue}
\citep[hereafter \kic,][]{Brown_2011} against \lamost\ Data Release (DR)
3 \citep{lamost} using a 1\arcsec\ search radius.  This match
revealed 53,090 sources. We cross matched the resulting sample
against the \allwise\ catalogue \citep{Wright_2010} using the \irsa\
web service and an increased search radius of 2\arcsec\ to account for
the larger point spread function in \allwise.  This revealed 48,999
unique sources.  We corrected bright \wise\ photometry using the 
prescription of \citet{Patel_2014}, and we discarded stars based on 
a number of quality criteria. First, we required that all stars have 
reported magnitudes in $g$, $i$, \ddofo, $W1$ and $W2$.  We further 
removed stars where there was uncertainty in the \wise\ photometry: 
$\sigma(W1) > 0.025$~mag, or $\sigma(W2) > 0.022$~mag, or if the 
$\chi^2$ value in the $W1$ or $W2$ profile fitting exceeded 2.  
We made no quality cuts based on spectroscopy (e.g., using any
information from \lamost). The distilled sample contains 25,668 stars.

%We first cross matched the \textit{Kepler Input Catalogue}
%\citep[hereafter \kic,][]{Brown_2011} against \apogee\ Data Release (DR)
%14 \citep{sdss_dr14} using a 1\arcsec\ search radius.  This match
%revealed 20,124 sources.  We cross matched the resulting sample
%against the \allwise\ catalogue \citep{Wright_2010} using the \irsa\
%web service and an increased search radius of 2\arcsec\ to account for
%the larger point spread function in \allwise.  This revealed 19,954
%unique sources.  We corrected bright \wise\ photometry using the 
%prescription of \citet{Patel_2014}.
%We discarded stars based on a number of quality criteria.
%First, we required that all stars have reported magnitudes in $g$, $i$,
%\ddofo, $W1$ and $W2$.  We further removed stars where there
%was uncertainty in the \wise\ photometry: $\sigma(W1) > 0.025$~mag, or
%$\sigma(W2) > 0.022$~mag, or if the $\chi^2$ value in the $W1$ or $W2$
%profile fitting exceeded 2.  
%

%We did not apply any quality cuts using 
%\apogee\ parameters. The distilled sample comprises 13,555 stars.

\begin{figure*}
  \begin{center}
    \hspace{-0.5cm} \includegraphics[width=1.0\textwidth]{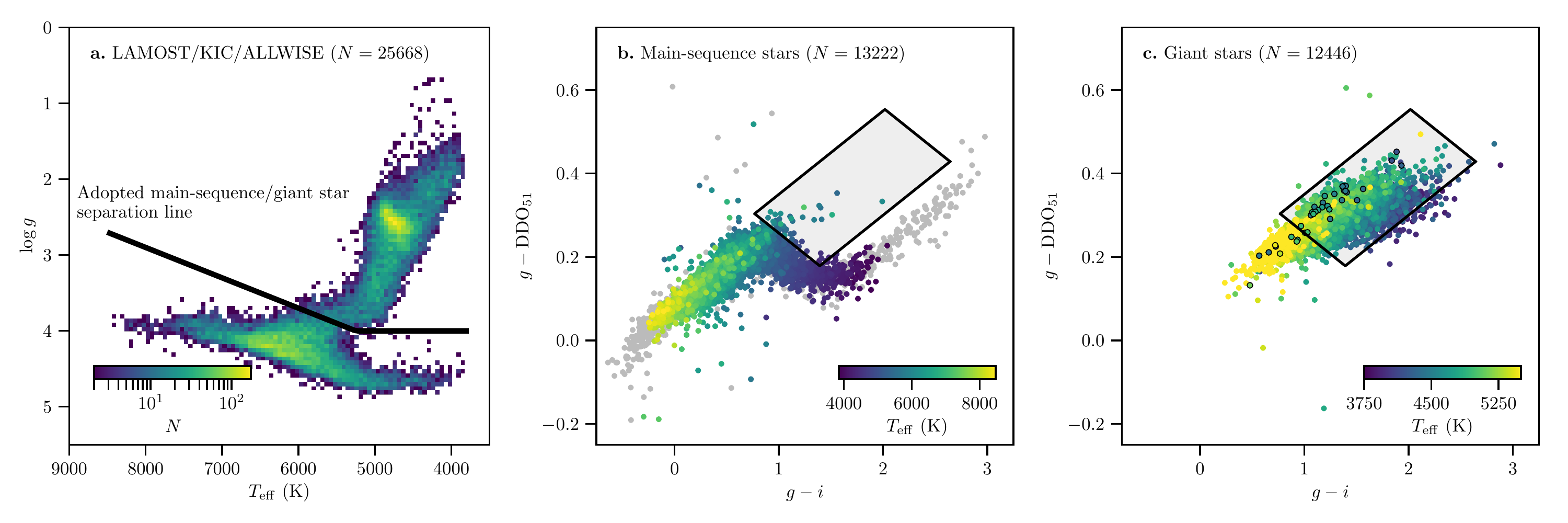}
    \caption{\textbf{a.} Spectroscopically-derived stellar parameters for
    stars in the \lamost/\kic/\allwise\ sample, with the line we
    adopted to separate main-sequence stars from giant stars. \textbf{b.}
    The $g-i$ and $g-$\ddofo\ colours of (spectroscopically-selected) 
    main-sequence stars in this \lamost/\kic/\allwise\ sample. The criteria
    box we adopt to photometrically select giant stars is shown in panel (b)  and (c).
    In (b) we also show main-sequence  stars in \apogee\ sample to demonstrate
    how redder main-sequence stars (which are not  present in \lamost) appear in this colour plane.
    \textbf{c.} $g-i$ and $g-$\ddofo\ colours of spectroscopically-selected giant stars in \lamost/\kic/\allwise, coloured (and ordered on the  plot) by their effective temperature. Stars with $[{\rm  Fe/H}] < -1.5$ are drawn with black edges, showing a mild metallicity dependence in the $g-$\ddofo\ colour. Most giants fall within the photometric selection, although the completeness drops for stars $T_{\rm eff} \gtrsim 5250\,{\rm K}$ as they overlap with main-sequence star colours in this plane.}\label{fig:sel}
  \end{center}
\end{figure*}

%\begin{figure}
%  \begin{center}
%    \hspace{-0.5cm} \includegraphics[width=0.5\textwidth]{gi-gd51.eps}
%    \caption{Selection of giants in the $g-D51$ vs. $g-i$ plane.  Each star is 
%    coloured by \apogee\ effective temperature.  Black points indicate no
%    effective temperature is reported by \apogee\ because the star
%    has been classified as a likely dwarf. Stars with ${\rm [M/H]} < -1.4$ 
%    are shown as larger dots.  Most giant stars lie above the selection 
%    function shown by the solid line.}\label{fig:sel}
%  \end{center}
%\end{figure}

%For the purposes of illustrating our photometric selection to separate
%dwarf and giant stars, we show the effective temperature $T_{\rm eff}$
%and surface gravity $\log{g}$ for the \lamost\ sample in the first 
%panel of Figure \ref{fig:sel}.  We classify all stars with
%$\log{g} > \max{\{6.1-2.4(T_{\rm eff}/6000), 4.1\}}$ as main-sequence
%stars. 

We use $g-$\ddofo\ colour to separate dwarf and giant stars, as shown
in Figure \ref{fig:sel}. In the first panel we show the effective
temperature $T_{\rm eff}$ and surface gravity $\log{g}$ for all \lamost\
stars in our sample, where we have separated main-sequence and giant
stars with $\log{g} > \max{\{6.1-2.4(T_{\rm eff}/6000), 4.1\}}$. The
$g-i$ and $g-$\ddofo\ colours of the main-sequence and giant star samples
are shown in the second and third panels of Figure \ref{fig:sel}, where
it is clear that the lack of spectral absorption in the giant
stars separates them very neatly from the dwarfs in the $\{g-i,g-$\ddofo$\}$
colour space, as shown in previous
studies.  We note that the separation we find in $g-$\ddofo\ appears
qualitatively better than existing studies using this selection, presumably
because these stars are relatively bright and the \ddofo\ imaging obtained
for the \kic\ is of high quality.  Using the $g-i$ colour, it is clear
that the dwarf/giant separation is maintained for effective temperatures
as low as $\sim\!\!4000$~K.  The separation for cooler stars is less distinct
for $g-r$, $J$, $H$, and $K_s$, but comparable for $g-z$.  For stars
below $\sim\!\!4000$~K, $V-K$ and $J-H$ can be effectively used to separate
dwarfs from giants \citep{Bessell_Brett_1988}.  On the hotter end, however,
we caution that $\log{g}$ sensitivity is largely lost for stars with 
$T_{\rm eff} \gtrsim 5250~{\rm K}$. Giant stars in this temperature regime
overlap with main-sequence stars (of all metallicities) and cannot be 
efficiently selected using only $g$, $i$, and \ddofo\ magnitudes without
introducing considerable contamination by main-sequence stars. For this 
reason, we can expect that any effective dwarf/giant selection using the 
$g-$\ddofo colour will be biased against hotter stars near the base of the
red giant branch.

% For the one very metal-poor giant
%star in our \apogee\ sample with $T_{\rm eff} \approx 5400~{\rm K}$, 
%Figure~\ref{fig:mh} suggests that giant stars with 
%$T_{\rm eff} \gtrsim 5300~{\rm K}$ maintain a relationship between 
%$W_1-W_2$ and [Fe/H], but they cannot be efficiently selected as giant 
%stars using the $g-$\ddofo\ colour, without substantial
%contamination by metal-rich dwarfs.

%In Figure \ref{fig:sel2} we show the $g-i$ and $g-$\ddofo\ colour for the
%full sample of  stars, coloured by \apogee\ metallicities where available.
%Note that the drawing of the points are ordered such that metal-poor stars
%appear on top of all others. Figure  \ref{fig:sel2} shows that the
%$g-$\ddofo colour is itself sensitive to metallicity \citep[as a second-order
%effect][]{Paltaoglou_1994,Majewski_2000}. Since our focus here is to cleanly
%select relatively metal-poor stars, we select giant stars that satisfy the
%following criteria, which is also shown in Figures \ref{fig:sel} and 
%\ref{fig:sel2}:

%\todo{UPDATE EQUATIONS}
%\begin{equation}\label{eq:sel}
%    g-{\rm DDO}_{51} > \max\left\{ \begin{array}{ll}
%    0.32 - 0.1 \times (g-i) \\
%    -0.1 + 0.2 \times (g-i) \, .
%    \end{array} \right.
%\end{equation}

The photometric selection we adopt for giant stars in this work is,
\begin{eqnarray}\label{eq:sel}
   g-{\rm DDO}_{51} &>& \max{\{0.46 - 0.2(g-i), -0.10 + 0.2(g-i)\}} \nonumber\\
   {\rm and}\nonumber\\
   g-{\rm DDO}_{51} &<& \min{\{0.15 + 0.2(g-i), +0.96 - 0.2(g-i)\}} . \nonumber
\end{eqnarray}

Using this selection we classify 8,947 stars as likely giants from the
\lamost/\kic/\allwise\ catalogue (of the 25,668 that met our photometric 
quality cuts).  Of these, 8,891 (99.4\%) are spectroscopically-confirmed 
giant stars, giving a contamination rate of main-sequence stars of 0.6\%. 
The completeness fraction is 71\%.
While the photometric selection could be adjusted to improve completeness,
here we have chosen simple criteria to maintain a low contamination
fraction.  Note that the completeness fraction is temperature-dependent, and
drops quickly for higher effective temperatures near $T_{\rm eff} \gtrsim 5250\,{\rm K}$,
as main-sequence stars and hotter giant stars share similar $g-i$
and $g-$\ddofo colours. The completeness fraction for \lamost\ giant 
stars with $T_{\rm eff} < 5250\,{\rm K}$ is 76\%. We stress, however, 
that the contamination and completeness 
fraction is subject to the \lamost\ selection function (e.g., magnitude
range, biases towards or against spectral types), and by the quality 
constraints that we have enforced on the \allwise\ photometry. 

We repeated the steps above using the \apogee\ DR14 \citep{sdss_dr14} 
to investigate the impact on completeness and contamination.
We find 20,124 matches between \apogee\ and the
\kic, and 19,952 unique sources that also have \allwise\ photometry.
After applying the same quality cuts described above, our distilled
\apogee/\kic/\allwise\ sample contained 13,555 stars. 
% Excluded APOGEE stars STARFLAG  = 4 or 131072
 Most of these stars are giants, a reflection of the \apogee\ selection bias towards
giant stars \citep{Zasowski_2013}. 
  Despite this strong bias in
favour of observing giant stars, we find that the estimated completeness
and contamination fraction arising from our photometric selection do not
change considerably. The completeness fraction is about 80\% -- whether
or not we restrict the sample to $T_{\rm eff} > 5250\,{\rm K}$ -- and
the contamination fraction is 0.6\%.  Since \apogee\ is biased towards
giant stars, this completeness fraction may be representative of an upper
limit of what could be expected from our colour selection, unless we 
were to adjust it for the sake of increasing contamination.
Nevertheless, although these estimated rates are sample-dependent, it is clear that
the \ddofo\ filter can be used to identify giant stars with very little
contamination whilst maintaining a reasonably
high completeness fraction. 
From the
\apogee\ and \lamost\ samples investigated, both showed comparable
completeness and contamination rates: the photometric selection for
giant stars recovered about 75\% of the true number of giants (as
determined by spectroscopy), and the contamination of main-sequence
stars is $\lesssim$1\%.

\begin{figure}
  \begin{center}
    \hspace{-0.5cm} \includegraphics[width=0.5\textwidth]{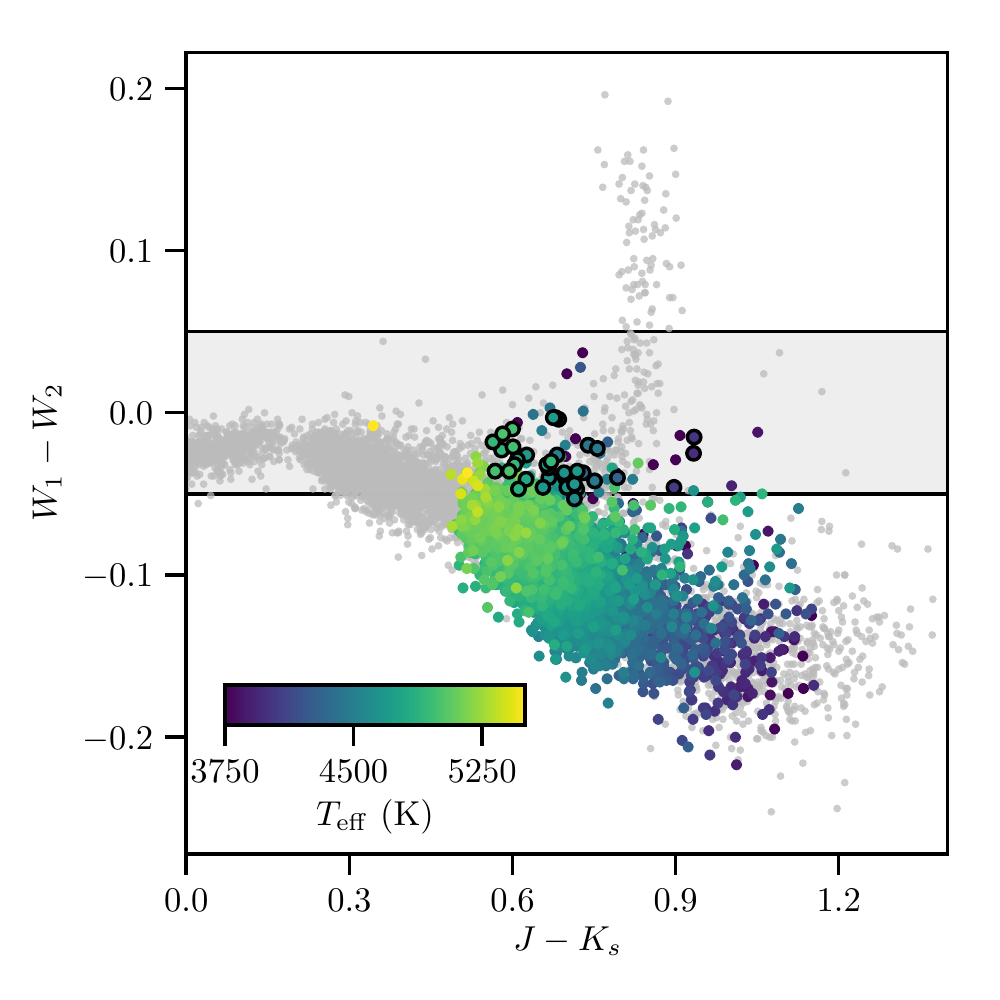}
    \caption{$J-K_s$ colours \citep{Skrutskie_2006} against $W1 - W2$
    colours for the \apogee/\kic/\allwise\ sample. Spectroscopically-confirmed
    giant stars are coloured by their effective temperature $T_{\rm eff}$ \citep{sdss_dr14},
    and all other points are shown in gray. Metal-poor giant stars ($[{\rm Fe/H}] < -1.5$)
    are marked with black edges, showing that all metal-poor giant stars in this sample
    have $-0.05 \lesssim W1-W2 \lesssim 0.05$ (indicated by the coloured region). 
    A photometric selection using only $W1-W2$
    colour would be severely contaminated by main-sequence stars (shown in grey).}
    \label{fig:infrared_colours}
  \end{center}
\end{figure}

%\emph{Top panel:} Metallicity of the giant sample as a function of
%    $W1-W2$ colour.  Stars are marked as in Figures \ref{fig:sel} and 
%    \ref{fig:corr}. \emph{Bottom panel:} Fraction of metal poor 
%    (${\rm [M/H]} < -1.4$) giants.  The dashed line shows the fraction of the 
%    total 8,008-star sample in each bin, while the solid line shows the fraction
%    of all stars in that $W1-W2$ bin (including both those at higher 
%    metallicities and with no \apogee-derived metallicities).

Using the \apogee\ sample, in Figure \ref{fig:infrared_colours} we plot \wise\ $W1-W2$
colour as a function of $J-K_s$, showing the same bifurcation in dwarf and giant stars noted by
\citet{Bessell_Brett_1988} for purely near-infrared colours. From Figure 
\ref{fig:infrared_colours} it can also be seen
that all metal-poor giant stars have a \wise\ $W1-W2$ colour of $\gtrsim -0.05$,
with a negligible temperature dependence.  The most metal-poor stars can be
retained by keeping giant stars with the reddest \wise\ $W1-W2$ colour.
However, the vast majority of giants in our sample are sufficiently warm that a
purely near-infrared selection of metal-poor giant stars would suffer heavy 
contamination by main-sequence stars. Coupled with our $g-$\ddofo photometric selection to
cleanly distinguish main-sequence stars from giant stars, we can now illustrate how this
sample permits the easy identification of metal-poor giant stars using
only photometry.

% Given that \wise\ colours
%are a function of stellar parameters (which may hinder our ability to
%robustly identify metal-poor stars), we first investigate whether a
%temperature correction is warranted.  Figure \ref{fig:corr} shows the
%$W1-W2$ dependence on effective temperature $T_{\rm eff}$, and a simple
%second-order polynomial fit (with coefficients 0.1, $-0.42$, 0.18) which
%could be applied to ensure $W1-W2$ is independent of stellar temperature.
%The dispersion of the $W1-W2$ colours of the corrected giant population
%increases towards cooler stars, which we attribute to reddening -- applying
%an analogous correction using spectroscopic temperatures yields a much
%tighter correlation.

%The obvious grouping of metal-poor stars in Figure \ref{fig:corr} shows
%that in the absence of accurate (e.g., spectroscopic) temperatures,
%such a correction is unwarranted.  We note that the low outlier
%point in Figure \ref{fig:corr} was incorrectly reported by \apogee\ to
%be metal-poor (${\rm [M/H]} \sim -1.9$).  The \texttt{ASPCAPFLAG = 0}
%flag indicates no suspicious heuristics in the analysis, although the
%\texttt{STARFLAG} column indicated some pixels with high persistence.
%After visual inspection of the spectrum, we conclude that this star is
%not a bonafide metal-poor star.  That is, in this case \wise\ photometry
%provided a simple way to reject a star that would otherwise be classified
%as very metal-poor.  

\begin{figure}
  \begin{center}
    \hspace{-0.5cm} \includegraphics[width=0.5\textwidth]{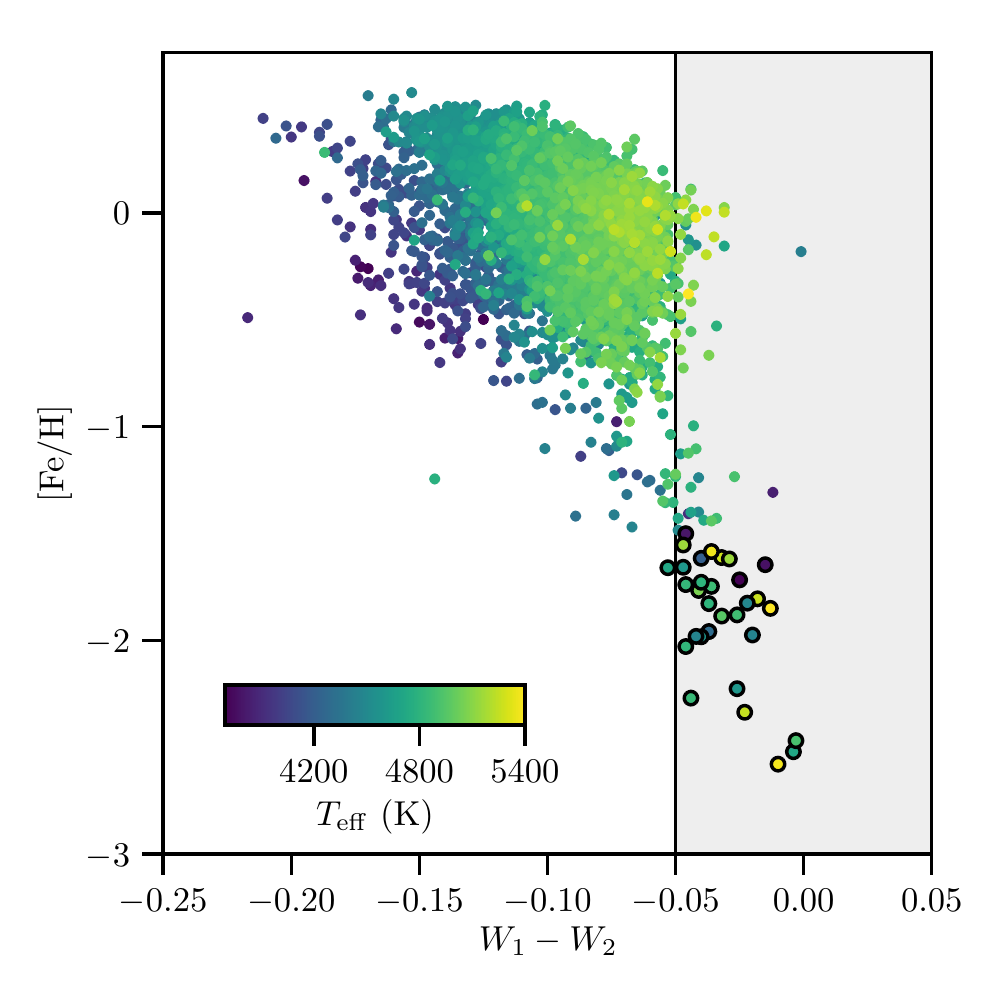}
    \caption{Metallicity of the \apogee/\kic/\allwise\ photometrically-selected
    giant sample ($N = 7,432$) as a function of $W1-W2$ colour. Points are 
    coloured by their \apogee\ effective temperature. Another 487 (6\%) photometrically-selected giant stars are not shown because they do not have metallicities in \apogee, likely due to analysis issues (see text).}
    \label{fig:infrared_mh}
  \end{center}
\end{figure}

Continuing with the \apogee\ sample, in Figure \ref{fig:infrared_mh} we show the
relation between \wise\ colour and \apogee\ metallicities
for 7,432 stars that meet our photometric giant star selection. There are another
487 (6\%) stars that we identify as likely giant stars (from photometry),
but \apogee\ does not report a metallicity.  Metallicity may not be reported
for a number of reasons, including: problems related to the data reduction
and/or analysis, or if \apogee\ suspects that the source is a main-sequence star.
Given the low contamination fraction we find from \lamost\ (e.g., $\lesssim$1\%), we 
suspect most of these 487 photometrically-selected giant stars do not have
reported metallicities due to the data reduction or analysis issues.
We checked those 487 photometrically-selected giant stars without \apogee\
metallicities and found that 158 have stellar parameters reported by \lamost,
and nearly all (152/158) are giant stars according to \lamost.
Based on the $W1 - W2$ colour separation of metal-poor giant stars ($[{\rm Fe/H}] < -1.5$) 
visible in Figures \ref{fig:infrared_colours} and \ref{fig:infrared_mh}, we adopt
$-0.05 \leq W1 - W2 \leq 0.05$, in conjunction with our $g-$\ddofo\ colour
selection, to distill a relatively clean sample  of metal-poor
giant  stars.

%When selecting metal-poor stars with just red \wise\
%colours, dwarf stars are the primary contaminants.  Roughly half of these
%contaminating stars can be removed with a more stringent giant selection,
%by increasing the intercept of the positive-slope cut in equation
%\ref{eq:sel} from $-0.1$ to 0.02.  This lower contamination rate however
%comes at the cost of a lower overall giant star completeness (i.e., some
%metal-rich giant stars are discarded), leaving only $\sim$5,000 giants.

\begin{figure*}
  \begin{center}
    \hspace{-0.5cm} \includegraphics[width=\textwidth]{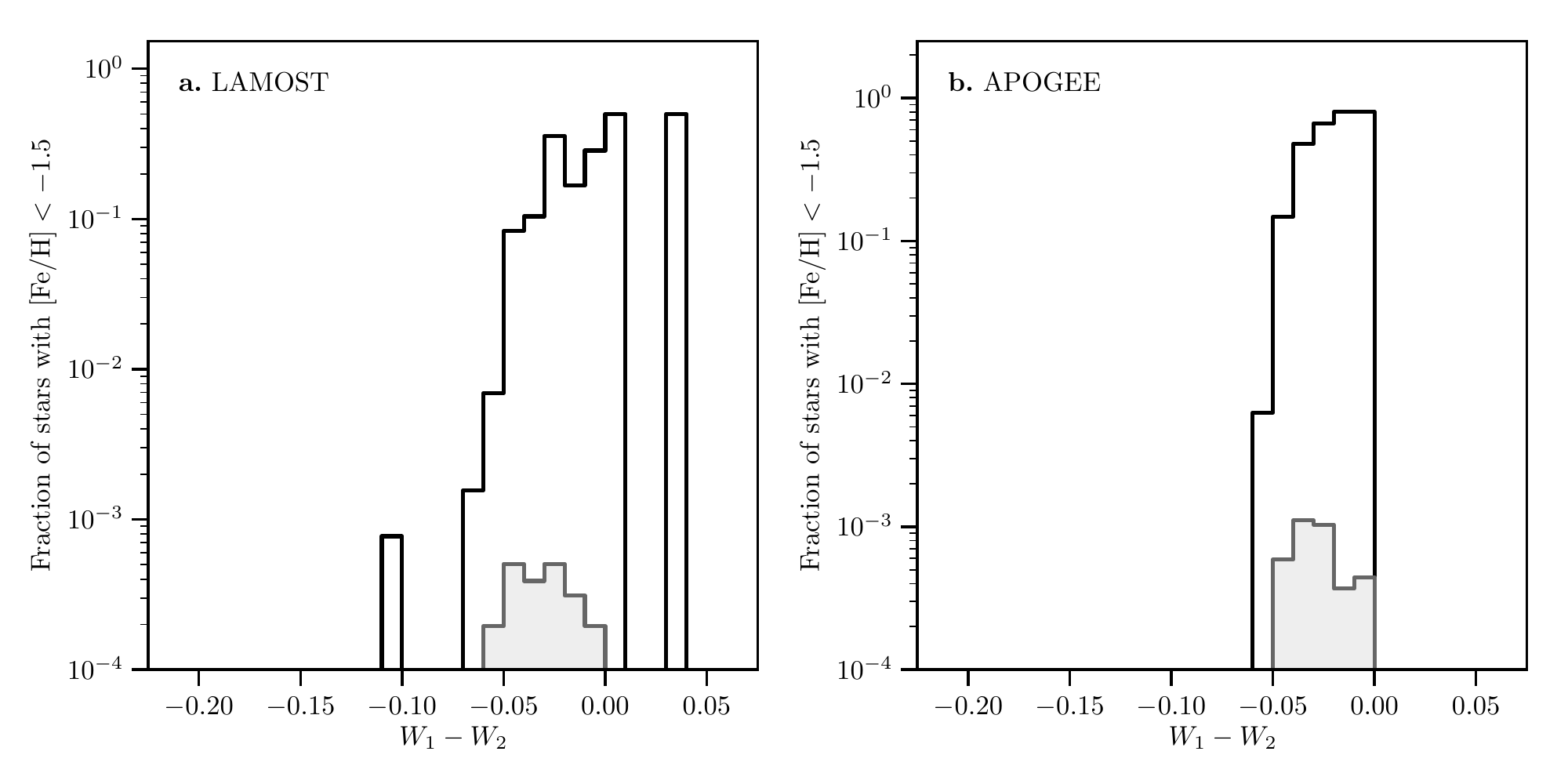}
    \caption{Effectiveness in photometrically selecting metal-poor ($[{\rm Fe/H}] < -1.5$) giant stars from $g-$\ddofo\ and $W1-W2$ coloured (black) over a random sample (grey). The lines indicate the fraction of metal-poor stars per $W1-W2$ colour bin. The use of the $g-$\ddofo\ colour selection removes most  metal-rich contaminants for colour bins at $W1 - W2 \geq -0.05$.}
    \label{fig:performance}
  \end{center}
\end{figure*}

Figure \ref{fig:performance} demonstrates the effectiveness of using the
$g-$\ddofo\ filter and \wise\ colours  to identify metal-poor
giant star candidates.  The greyed area shows the fraction of metal-poor
giant stars ($[{\rm Fe/H}] < -1.5$) per $W1 - W2$ colour bin for the 
\lamost\ and \apogee\ samples.  Given a single photometric selection in
$W1 - W2 > -0.05$, without any giant/dwarf selection, the fraction of
metal-poor stars is $\sim\!\!1:1,000$ stars per bin.
In Figure \ref{fig:performance} we also show the fraction of metal-poor
stars per colour bin for our photometrically-selected sample of giants.
Above $W1 - W2 > -0.05$, the fraction of stars with $[{\rm Fe/H}] < -1.5$
exceeds 25\%. Thus, our photometric selection has increased the yield of
metal-poor giant stars recovered by a factor of $\sim$250 over a
randomly selected sample.

% We also show with a solid line the
%fraction of metal-poor stars that fall in each $W1-W2$ bin, calculated by
%the number of stars with ${\rm [M/H]} < -1.4$ divided by the total number
%of stars that survived the giant photometric selection.  Above $W1-W2 >
%-0.05$, the fraction of stars with ${\rm [M/H]} < -1.4$ exceeds 24\%.
%Thus, our photometric selection has increased the yield of metal-poor
%giant stars recovered by a factor of $\sim$250 over a randomly selected
%sample.

\subsection{Weighing the Giants}
\label{sec:weighing-the-giants}

We have presented an efficient and effective photometric selection 
to identify giant metal-poor stars.  Given the paucity of
known metal-poor stars in the \kepler\ field and the routine use of
asteroseismic scaling relations to estimate stellar masses and radii,
this provides us with a unique opportunity to critically examine the
existing scaling relations.  While the number of metal-poor giant
stars with publicly-available asteroseismic fundamental parameters
($\nu_{\rm max}$, $\Delta\nu$) is very small, there is theoretical
and observational evidence to suggest that the current scaling
relations warrant a metallicity-dependent term in order to reconcile
systematically over-estimated masses inferred from metal-poor stars
\citep[e.g.,][]{White_2011,Mosser_2013,Epstein_2014,Gaulme_2016,
Guggenberger_2016,Sharma_2016,Huber_2017}.  Here we cross match
our catalog of photometrically-selected metal-poor stars with publicly
available asteroseismic fundamental parameters to infer their masses and
radii, and explore whether any correction is necessary to the existing
scaling relations.

We cross matched the \kic\ and \allwise\ and selected likely metal-poor
giant stars based on $g-$\ddofo\ and $W1-W2$ colours, while maintaining
the photometric quality cuts described earlier. We cross matched the resulting
sample against the \citet{Hekker_2011} catalog, which includes a thorough 
comparison of six different asteroseismic
pipelines.  Our search revealed four new highly likely metal-poor 
($[{\rm Fe/H}] \lesssim -1.5$) giant
stars with measured $\Delta\nu$ and $\nu_{\rm max}$ from multiple 
pipelines: \kic{}s 6304081 (all 6 pipelines), 6231193 (6), 7729396
(3), and a fourth which will appear in a subsequent paper (Schlaufman
et al. 2018, in preparation).  The pipeline measurements are typically
in agreement within a few percent for all three stars (see Table
\ref{tab:oscillation-parameters}).

We assume $T_{{\rm eff},\odot} = 5777$~K for the Sun, and adopt
$\Delta\nu_\odot = 135.0 \pm 0.1$~$\mu$Hz and $\nu_{{\rm max},\odot}
= 3140 \pm 30$~$\mu$Hz as per \citet{Epstein_2014}.  If we employ the
photometric temperatures from the \kic\ and take the mean $\nu_{\rm max}$
and $\Delta\nu$ values from Table \ref{tab:oscillation-parameters},
the asteroseismic scaling relations imply respective solar masses of
1.01, 2.19, 1.04, and solar radii of 10.7, 10.6, 20.5 for our metal-poor
giant star candidates. The estimated uncertainties on these masses is
of order 10\% \citep[e.g.,][]{chaplin_2013}.
The indicated masses are relatively high for 
metal-poor stars. Metal-poor stars are generally thought to be ancient,
which demands that they should have masses of $\approx$0.8$M_\odot$ 
in order to survive to the present day. Masses higher than
$\approx$0.8$M_\odot$ for ancient stars suggest that they would live
relatively short lifetimes and not remain observable today. Given the good 
agreement in global oscillation parameters reported from multiple 
pipelines, the choice of mean oscillation parameters or those from any individual pipeline has little
impact on the inferred masses and radii. For example, the scatter in 
oscillation parameters for KIC 6231193 translates to a scatter in 
inferred mass of just  $\sigma(M) = 0.03\,M_\odot$. We also note the effective
temperatures estimated by the \kic\ catalog agree excellently with the
distribution of temperatures found for confirmed metal-poor giant stars in
the \citet{Schlaufman_Casey_2014} catalog.  The photometric temperatures
would have to be systematically overestimated by $\sim$2,500~K to bring
all inferred masses within 0.8~$M_\odot$. 

% KIC     % <Nu_max> \pm \sigma(Nu_max) & <Dnu> \pm \sigma(Dnu)
% 6231193 & $31.37 \pm 1.05$ & $3.89 \pm 0.07$ \\
% 6304081 & $66.92 \pm 2.79$ & $5.83 \pm 0.22$ \\
% 7693833 & $31.99 \pm 0.88$ & $4.05 \pm 0.11$ \\
% 7729396 & $8.80 \pm 0.43$ & $1.48 \pm 0.12$ \\

% KIC   M_sun R_sun
% --> By taking mean Nu_max, Dnu.
%6231193 1.01 10.68
%6304081 2.19 10.55
%7693833 1.00 10.37
%7729396 1.04 20.53

\begin{table*}
\begin{tabular}{lccccccccccccccccccc}
\hline
& \multicolumn{2}{c}{COR} && \multicolumn{2}{c}{CAN} && \multicolumn{2}{c}{A2Z} && \multicolumn{2}{c}{SYD} && \multicolumn{2}{c}{DLB} && \multicolumn{3}{c}{OCT} \\
 \cline{2-3} \cline{5-6} \cline{8-9} \cline{11-12} \cline{14-15} \cline{17-19}
\kic\ & $\nu_{\rm max}$ & $\Delta\nu$ &&  $\nu_{\rm max}$ & $\Delta\nu$ && $\nu_{\rm max}$ & $\Delta\nu$ && $\nu_{\rm max}$ & $\Delta\nu$ && $\nu_{\rm max}$ & $\Delta\nu$ && $\nu_{\rm max,1}$ & $\nu_{\rm max,2}$ & $\Delta\nu$ \\
& $\mu$Hz & $\mu$Hz && $\mu$Hz & $\mu$Hz && $\mu$Hz & $\mu$Hz && $\mu$Hz & $\mu$Hz && $\mu$Hz & $\mu$Hz && $\mu$Hz & $\mu$Hz & $\mu$Hz \\
\hline
6231193 & 31.5 & 3.93 && 31.22 & 3.91 && 31.4  & 3.92 && 31.84 & 3.93 && 29.09 & 3.74 && 31.72 & 32.84 & 3.89 \\
6304081 & 69.1 & 5.78 && 68.57 & 5.48 && 67.03 & 5.79 && 69.36 & 5.9  && 60.65 & 6.23 && 67.75 & 65.95 & 5.79 \\
%7693833 & 31.3 & 4.19 && 31.76 & 4.01 && 31.36 & 3.86 && 31.48 & 4.05 && \nd   & \nd  && 32.17 & 33.84 & 4.13 \\
7729396 & 8.3  & 1.41 && \nd   & \nd  && \nd   & \nd  && 8.76  & 1.39 && 9.35  & 1.65 && \nd   & \nd   & \nd  \\ 
\hline
\end{tabular}
\caption{Global oscillation parameters measured from six different pipelines for photometrically-selected metal-poor giant star candidates. Pipeline acronyms are as per \citet{Hekker_2011}.}
\label{tab:oscillation-parameters}
\end{table*}

\section{Discussion}
\label{sec:discussion}

We have leveraged existing public photometry and spectroscopy in
the \kepler\ field to verify that the \ddofo\ filter is successful
in separating FGK dwarf and giant stars.  While late-type (${\rm
M}0.0$ and later) giant stars are difficult to separate from dwarf
stars given the $g-$\ddofo\ colour, infrared photometric selections
already exist that allow for a clean selection of M-type giant stars
\citep{Bessell_Brett_1988}.  Our photometric selection for giants is
extremely simple (i.e., we did not choose to optimise parameters to
maximise yields), and we estimate the contamination in our giant sample
to be $\sim$1\%, subject to the selection function of the \apogee\ 
and \lamost\ samples used. The completeness fraction is estimated to 
be about 75\%, with a bias against giant stars with $T_{\rm eff} \gtrsim 5150\,{\rm K}$.

The dwarf/giant separation power of \ddofo\ is well established, and
there is no doubt it was the primary reason that the \kic\ was successful
in identifying dwarf stars \citep{Verner_2011}.  Here we have shown that
the giant stars in the resulting sample have spectroscopic
metallicities that are strongly correlated with $W1-W2$ infrared colour.
This correlation is not present in dwarf stars at a level that permits
metal-poor stars to be easily identified.  Indeed, Figure \ref{fig:infrared_colours}
demonstrates that dwarf stars of all metallicities display infrared
colours that are indistinguishable from metal-poor giant stars.  For this
reason, a dwarf/giant discriminant is required to reveal the correlated
signature between stellar metallicity and infrared colour.

An examination of theoretical stellar spectra reveals the explanation for
this strong correlation between metallicity and $W1-W2$ colour.  Using the
PHOENIX spectral library \citep{Husser_2013}, \citet{Kennedy_Wyatt_2012}
and \citet{Schlaufman_Casey_2014} showed that the dependence of
metallicity on $W1-W2$ relies on the presence of strong CO absorption
at $\approx4.5$~$\mu$ (i.e., in the middle of $W2$).  This band-head is
strong, and only begins to disappear at ${\rm [Fe/H]} \lesssim -2$ for
a giant star with $T_{\rm eff} \approx 4800$~K.  There are no strong
stellar absorption features in wavelengths covering the $W1$ band,
therefore making $W1-W2$ a sensitive proxy for stellar metallicity.
%While the correlation between stellar 
%metallicity and infrared colour corroborates expectations from theoretical
%stellar spectra, the clarity of the relationship is impressive.

\citet{Schlaufman_Casey_2014} first utilised $W1-W2$ colour to
efficiently identify bright metal-poor stars from existing public data.
Their selection is as efficient as existing studies seeking metal-poor
stars, while the candidates they identify are several magnitudes brighter.
This property minimises the requisite telescope time for spectroscopic
confirmation and subsequent detailed chemical abundance analysis.
A number of photometric cuts are employed by \citet{Schlaufman_Casey_2014}
in addition to the $W1-W2$ colour.  Some cuts accounted for expected
temperature dependencies, while others were included because they
empirically improved the yield of metal-poor giant stars.

Young, relatively metal-rich dwarf stars are the primary contaminant
that result from the original \citet{Schlaufman_Casey_2014} photometric
selection.  Our analysis of public photometry and spectroscopy in the
\kepler\ field has revealed the reason for this contamination and
further verified the potential in using infrared colours to select
metal-poor giant stars.  When coupled with a simple dwarf/giant
photometric selection, our analysis shows that the $W1-W2$ infrared
sensitivity alone is sufficient to effectively select metal-poor stars.
It would appear the additional empirical photometric cuts employed by
\citet{Schlaufman_Casey_2014} primarily served to minimise the dwarf
contamination, thereby improving the overall yield of metal-poor giant
stars.

Infrared photometric selections are advantageous because they are
minimally affected by strong absolute or differential extinction.
While we have shown the metallicity sensitivity in giants is
principally correlated with the mid-infrared $W1-W2$ colour, the
central wavelengths of $g$ and \ddofo\ nearly overlap at 510~nm.
Due to the their bluer wavelengths, one might expect the $g$ and \ddofo\
filters to be considerably impacted by extinction.  However, because
the $g$ and \ddofo\ filters have central wavelengths that are very
similar, both filters are affected by dust in a similar way.  For this
reason, the $g-$\ddofo\ colour has a very small dependence on extinction.
Therefore, a photometric selection of metal-poor stars that makes use
of $g-$\ddofo\ and $W1-W2$ will be minimally impacted by dust.  This is
particularly relevant for metal-poor star searches in the inner Galaxy,
where absolute and differential extinction is strongest.  This is
important, as theoretical models of galaxy formation predict that the
oldest stars in the Milky Way should be metal-poor stars found in the
bulge \citep[e.g.,][]{Tumlinson_2010}.

We have employed our photometric selection to identify likely metal-poor
giant star candidates with publicly available global oscillation
parameters.  There is excellent agreement between reported parameters from
different pipelines for our candidates.  Standard asteroseismic scaling
relations imply masses exceeding 1~$M_\odot$, up to 2.19~$M_\odot$ for the
most extreme case.  For metal-poor giant stars -- which are presumably old
halo stars -- the expected masses are $\sim$0.8~$M_\odot$.  Higher masses
are inconsistent with the requisite ancient age of the halo stars.
Although our sample size is small ($N = 3$), this discrepancy implies
that the standard scaling relations would have to be over-predicting
the masses of metal-poor stars by 25\% to $\sim$175\% in order to be
consistent with the observations. Although there
are theoretical modifications to standard scaling relations that attempt
to correct for some of this discrepancy \citep[e.g.,][]{White_2011}, the impact is of the order 5\%.
Therefore, our metal-poor giant star candidates suggest that either a
stronger  metallicity-dependent correction is necessary to resolve this
discrepancy, or that metal-poor stars have much higher masses than
expected from stellar evolution.

\section{Conclusions}
\label{sec:conclusions}

We have derived a simple photometric selection for FGK-type giants
based on public $g$, $i$, and \ddofo\ photometry in the \kepler\ field.
Given a single two-colour cut, we find giant completeness rates of
$\sim$75\%, with $\lesssim$1\% contamination of dwarfs, subject to the
selection functions of \lamost\ and \apogee, and the magnitude ranges
considered.  We distill a sample
of photometrically-selected giant stars and show a strong correlation
between spectroscopic metallicity and mid-infrared $W1-W2$ colour.
This relationship is not present in dwarfs, so metal-poor candidate stars
selected solely from mid-infrared $W1-W2$ excess will be contaminated
by dwarf stars across all metallicities.

Our work extents that of \citet{Schlaufman_Casey_2014}, who first
showed that metal-poor stars could be successfully identified through
infrared colours \citep[see also][]{Kennedy_Wyatt_2012}.  Here we have
shown that the additional cuts used in \citet{Schlaufman_Casey_2014}
to empirically improve the yield of metal-poor giant stars primarily
act to minimise the number of dwarf stars in the sample, rather than
principally discriminating on metallicity.

Our photometric cuts are well-founded theoretically.  The $\log{g}$
sensitivity in $g-$\ddofo\ arises from pressure-sensitive spectral
features that are strong in dwarfs but weak or non-existent in giants of
the same temperature.  Similarly, the dependence of $W1-W2$ colour on
metallicity relies on negligible spectral features in $W1$, but strong
molecular CO absorption present in giant stars at $\approx4.5$~$\mu$
(i.e., in $W2$).  Here we have demonstrated that coupling these two
simple photometric selections provides enormous potential in robustly
identifying metal-poor stars, even in heavily extincted regions (e.g.,
the bulge).  For these reasons, we argue that a photometric selection
that employs $g-$\ddofo\ and $W1-W2$ colours to identify metal-poor
star candidates in the inner Galaxy may be the most promising way to
discover any remaining low-mass Population III in the Galaxy.

We have identified metal-poor giant star candidates in the \kepler\
field that have publicly available global oscillation parameters from
asteroseismic pipelines.  There is good agreement between different
asteroseismic pipelines for these stars.  However, the masses inferred
from standard scaling relations are higher than expectations for ancient
metal-poor stars.  These results imply
that either asteroseismic scaling relations over predict masses for
metal-poor giant stars, or that the metal-poor giant stars examined here
are younger than expected.

\section*{Acknowledgments}

We thank the anonymous referee for a prompt and detailed review.
We thank Gerry Gilmore, Mike Irwin, Melissa Ness, and Adrian Price-Whelan. 
This work was partly supported through the Australian Research Council 
through Discovery Grant DP160100637 and by the European Union FP7 programme 
through ERC grant numbers 320360 and 279973. GMK is supported by the 
Royal Society as a Royal Society University Research Fellow. T.~R.~H 
gratefully acknowledges support from an Undergraduate Research Bursary
from the Royal Astronomical Society.  This research made use of Astropy, 
a community-developed core Python package for Astronomy \citep{astropy}; 
NASA's Astrophysics Data System Bibliographic Services; and \texttt{TOPCAT}
\citep{topcat}. 

Kepler was competitively selected as the tenth Discovery 
mission.  Funding for this mission is provided by NASA’s Science Mission 
Directorate.

Funding for the Sloan Digital Sky Survey IV has been provided by the Alfred P. Sloan Foundation, the U.S. Department of Energy Office of Science, and the Participating Institutions. SDSS-IV acknowledges
support and resources from the Center for High-Performance Computing at
the University of Utah. The SDSS web site is www.sdss.org.
SDSS-IV is managed by the Astrophysical Research Consortium for the 
Participating Institutions of the SDSS Collaboration including the 
Brazilian Participation Group, the Carnegie Institution for Science, 
Carnegie Mellon University, the Chilean Participation Group, the French Participation Group, Harvard-Smithsonian Center for Astrophysics, 
Instituto de Astrof\'isica de Canarias, The Johns Hopkins University, 
Kavli Institute for the Physics and Mathematics of the Universe (IPMU) / 
University of Tokyo, Lawrence Berkeley National Laboratory, 
Leibniz Institut f\"ur Astrophysik Potsdam (AIP),  
Max-Planck-Institut f\"ur Astronomie (MPIA Heidelberg), 
Max-Planck-Institut f\"ur Astrophysik (MPA Garching), 
Max-Planck-Institut f\"ur Extraterrestrische Physik (MPE), 
National Astronomical Observatories of China, New Mexico State University, 
New York University, University of Notre Dame, 
Observat\'ario Nacional / MCTI, The Ohio State University, 
Pennsylvania State University, Shanghai Astronomical Observatory, 
United Kingdom Participation Group,
Universidad Nacional Aut\'onoma de M\'exico, University of Arizona, 
University of Colorado Boulder, University of Oxford, University of Portsmouth, 
University of Utah, University of Virginia, University of Washington, University of Wisconsin, 
Vanderbilt University, and Yale University.

Guoshoujing Telescope (the Large Sky Area Multi-Object Fiber Spectroscopic Telescope LAMOST) is a National Major Scientific Project built by the Chinese Academy of Sciences. Funding for the project has been provided by the National Development and Reform Commission. LAMOST is operated and managed by the National Astronomical Observatories, Chinese Academy of Sciences.
This publication makes use of data products from the Wide-field Infrared Survey Explorer, which is a joint project of the University of California, Los Angeles, and the Jet Propulsion Laboratory/California Institute of Technology, funded by the National Aeronautics and Space Administration.

\label{lastpage}

\end{document}